\documentclass[preprint]{aastex}
\usepackage{geometry}  
\usepackage{graphicx}
\usepackage{amssymb}
\usepackage{epstopdf}
\usepackage{hyperref}
\title{Limits on the Gas Disk Content of Two ``Evolved" T Tauri Stars}
\author{Alan G. Aversa}
\affil{Steward Observatory, 933 N Cherry Ave., Tucson AZ 85721}

\begin{abstract}
We derived upper limits of the circumstellar gas disk masses around the T~Tauri stars St~34 and RX~J0432.8+1735 in order to place constraints on theories of planet formation and to explore the evolution of the gas-to-dust ratio during the epoch of disk dissipation around young sun-like stars. Since sub-millimeter lines of $^{12}$CO trace of the cold, outer regions of circumstellar disks, we observed $^{12}$CO~$J=2-1$ emission with the 10~m Sub-Millimeter Telescope (SMT) for two carefully chosen targets. St~34 is a rare classical T Tauri star with an age of $8\pm3$~Myr, and RX~J0432.8+1735 is a rare weak-emission T Tauri star with far-infrared excess. Both exhibit radial space motion enabling us to distinguish disk emission from ambient cloud material. Assuming a $^{12}$CO excitation temperature of 20~K, a $^{12}$CO line-width of 5~km~s$^{-1}$, and optically-thin emission, we derive $3\sigma$ upper limits on the H$_2$ circumstellar disk mass for St~34 and RX~J0432.8+1735 to be $<4.20$~M$_\earth$ for both disks. Placing these results in the context of other studies, we discuss their implications on planet formation models.
\end{abstract}

\begin{document}
\maketitle

\section{Introduction}

Circumstellar disks of gas and dust, a natural result of the conservation of angular momentum, are a common outcome of the star formation process. \citet{kenyon95} find that over half the low mass ($<$ 2 M$_{\odot}$) pre--main-sequence T Tauri stars in the Taurus-Auriga star formation region have more infrared emission than expected from a normal stellar photosphere, indicating the presence of a dusty circumstellar disk heated by the parent star as well as active accretion. T Tauri stars fall into two categories: Weak-Line T Tauri Stars (WTTSs), characterized by low H$\alpha$ equivalent widths, and Classical T Tauri Stars (CTTSs), with higher H$\alpha$ equivalent widths indicative of ongoing gas accretion. These circumstellar disks generally have the following properties \citep{dutrey07,andrews09}: mass surface densities $\Sigma(r)\propto r^{0\,\mathrm{to}\,-1.0}$, surface temperatures $T(r)\propto r^{0\,\mathrm{to}\,-0.6}$(depending on disk flaring), and Keplerian rotational velocities $V(r)\propto r^{0.5}$.


Gas giant planet formation depends upon the gas content of the circumstellar disk from which they form. Primordial inner disks traced by hot dust disappear after about 3~Myr with a range of $1-10$~Myr \citep{meyer07}. If gas content and dust content in disks dissipate through similar mechanisms \citep{damajanov07}, then we would expect gas to disappear on these same timescales. To understand the timescales of planet formation, we must understand how long a gas disk persists around its parent star. Accretion rates (higher in CTTSs than WTTSs) also trace the time evolution of gas content as gas must be present to accrete. \citet{durisen07} suggest that gravitational instabilities in gas disks could account for their rapid ($<3$~Myr) dissipation, locking up mass in planets. In contrast, simulations of the evolution of gas disk surface density for a 1~M$_\sun$ star due to photo-evaporation indicate that gas disks disappear in about 6~Myr \citep{alexander06,dullemond07}. Because there are various theoretical results for the gas dispersal mechanisms and associated timescales, we might expect to observe diverse properties for gas disks around T Tauri stars.


Ideally, sub-millimeter interferometric images of T~Tauri stars would yield the most information about circumstellar disks, including their orientation, geometry, and gas content. Yet observations of this sort have only been published for a few nearby stars \citep[e.g., Sub-Millimeter Array (SMA) observations of TW~Hya;][]{qi04}. Infrared photometric campaigns with {\it Spitzer}, such the Cores to Disks \citep{evans03} and the Formation and Evolution of Planetary Systems \citep[FEPS;][]{meyer06} {\it Spitzer} Legacy projects, provide knowledge of the dust circumstellar disks based upon IR excesses at many wavelengths including $24\micron$, which trace dust within a few AU of the parent star, and at $70\micron$, which trace cooler dust at larger radii. \citet{silverstone06} surveyed 74 stars with ages between 3-30 Myr finding no stars with mid--IR excess that were not gas rich accreting T Tauri stars. \citet{padgett06} however identified a handful of WTTS (lacking signatures of accretion) with evidence for mid-- and far--IR excess emission. Photometric IR observations do not, however, constrain the total gas content in circumstellar disks because of a potentially highly variable gas-to-dust ratio. Observing the rotational energy transitions of $^{12}$CO, a proxy for molecular hydrogen (H$_2$), and assuming the ISM abundance ratio $\left[\mathrm{H}_2/\mathrm{CO}\right]\approx10^4$ enables one to trace the majority of cold gas in disks out to radii many times larger. Is the timescale for gas dissipation similar to the timescale for dust dissipation around T Tauri stars? This is the central question we address with our new observations. 


Expanding on previous works \citep[e.g.][]{pascucci06,vankempen07}, we search for $^{12}$CO~$J=2-1$ emission for two ``evolved" T~Tauri stars in Taurus to place constraints on their gas circumstellar disk masses. St~34 is evolved in the sense that it is $8\pm3$~Myr old, whereas RX~J0432.8+1735 is evolved in the sense that---although it is much younger ($1.0\pm0.5$~Myr)---it has already lost its inner accretion disk. In \S2, we describe our selection of sources and observations, then in \S3 we derive upper limits on our sources' gas disk masses. Lastly, in \S4 we describe the implications of our results, and in \S5 we list our conclusions.

\section{Observations}

\subsection{Selection of Sources}
We observed the $^{12}$CO~$J=2-1$ emission of the Taurus region objects St~34 and RX~J0432.8+1735, T Tauri stars with known 24 or $70\micron$ excesses \citep[see Table~\ref{sourcetab}]{kenyon95,padgett06}. In order to detect gas in their circumstellar disks, we must be able to distinguish emission from the disk and surrounding molecular cloud. From available candidate ``evolved'' (in age or shape of spectral energy distribution) T Tauri stars, we selected those least likely to be contamined by the ambient CO emission from the parent molecular cloud based on the \citet{dame87} CO survey. The radial velocities of our sources differ from the systemic velocities of any CO emission in the vicinity of our sources by $\sim2-3$~km~s$^{-1}$.

\begin{deluxetable}{lccccccccccll}
\tabletypesize{\scriptsize}
\tablewidth{0pt}
\rotate
\tablecaption{Candidate Source Summary\label{sourcetab}}
\tablehead{\colhead{Object} & \colhead{$\alpha$} & \colhead{$\delta$} & \colhead{Heliocentric RV} & \colhead{Source $V_\mathrm{LSR}$\tablenotemark{a}} & \colhead{Cloud $V_\mathrm{LSR}$\tablenotemark{b}} & \colhead{$T_\mathrm{eff}$} & \colhead{$\log{L^\star}$} & \colhead{Mass} & \colhead{Age} & \colhead{General} \\
\colhead{} & \colhead{(J2000)} & \colhead{(J2000)} & \colhead{(km~s$^{-1}$)} & \colhead{(km~s$^{-1}$)} & \colhead{(km~s$^{-1}$)} & \colhead{(K)} & \colhead{($\log{L_\sun}$)} & \colhead{($M_\sun$)} & \colhead{(Myr)} & \colhead{Reference}}
\startdata
RX~J0432.8+1735 & 04 32 53.23 & 17 35 33.68 & $18.6  $ & 7.3 & 10.2 & 3499\tablenotemark{c} & $\sim -6.1\times10^{-2}$\tablenotemark{g} & \nodata & $1\pm 0.5$\tablenotemark{d} & a\\
St~34  & 04 54 23.70 & 17 09 54.00 & $17.9 \pm 0.6 $ & 6.6 & 8.37 & 3415 & $1.03\pm0.06$ & $0.37\pm0.08$\tablenotemark{e} & $8\pm 3$\tablenotemark{e,f} & b
\enddata
\tablerefs{For RVs, see \citet{wichmann00,white05}. For general references, see (a) \citet{padgett06} and (b) \citet{white05}.}
\tablenotetext{a}{We corrected the heliocentric radial velocities (RVs) from the literature into local standard of rest ($V_\mathrm{LSR}$) radial velocities assuming the sun moves toward J2000 $(\alpha = 18\fh0, \delta = 30\fdg0)$ at 20~km~s$^{-1}$.}
\tablenotetext{b}{Determined by the $^{12}$CO~$J=1-0$ emission from the \citet{dame87} survey}
\tablenotetext{c}{\citet{wichmann00}}
\tablenotetext{d}{\citet{dantona97}}
\tablenotetext{e}{For both binary components}
\tablenotetext{f}{Isochronal age is given. The Li depletion age for both binary components is $>25$~Myr.}
\tablenotetext{g}{Estimated by integrating the spectral energy distribution (SED) of RX~J0432.8+1735 in \citet{padgett06}}
\end{deluxetable}


\subsubsection{St~34}
St~34 (HBC~425) is a binary system of two CTTSs, separated by $\lesssim0.78$~AU \citep{white05} based on the orbital solution for the system \citep{downes88} in the Taurus-Auriga T association \citep{kenyon95}. Both components of the spectroscopic binary have roughly equal mass and spectral types of M3 \citep{white05}. \citet{white05} observed St~34 in the optical with the HIRES spectrograph at Keck and derived an isochronal age of $8\pm3$~Myr for both components of the binary. Since they did not detect any lithium ($^7$Li) in the spectrum, St~34 must have reached---assuming the stars are completely convective---an internal temperature $>2\times10^6$~K and since depleted all of its lithium. St~34 has a low accretion rate of $2.5\times10^{-10}$~M$_\sun$~yr$^{-1}$, and the maximum radial velocity difference between the two binary components of St~34 is 58.4~km~s$^{-1}$ \citep{white05}. St~34, being one of the oldest known pre--main-sequence (PMS) star still accreting from a proto-planetary disk, also has a low dust mass of $\sim 2\times10^{-10}$~M$_\sun$ for radii $\lesssim 0.7$~AU \citep{hartmann05}.

\subsubsection{RX~J0432.8+1735}
RX~J0432.8+1735 is a WTTS of spectral type M2 \citep{martin99}. Based on the PMS tracks of \citet{dantona97} RX~J0432.8+1735 is estimated to be $1.0\pm0.5$~Myr old. \citet{padgett06} observed RX~J0432.8+1735 with {\it Spitzer} and noticed that its $24\micron$ flux is in excess of the expected photospheric value by a factor of 3. Its lack of IR excess $\le 12\micron$ suggests there may be a large inner hole in the disk. Based on {\it ROSAT} observations, \citet{carkner96} discovered that RX~J0432.8+1735 is also an X-ray source. As RX~J0432.8+1735 is classified as a WTTS star with no estimates of its accretion rate, we assume it is not accreting.

\subsection{Observing Procedure}
On 26-27 November 2007, we observed the $^{12}$CO~$J=2-1$ (230.53799~GHz) emission line of our two T~Tauri stars with the 10~m Heinrich Hertz Sub-Millimeter Telescope (HHT) on Mt.~Graham, Arizona. Observations were obtained with a 1~mm dual polarization (Vpol, Hpol), sideband-separating, ALMA prototype receiver. The upper sideband was tuned to $^{12}$CO~$J=2-1$ while the lower sideband was tuned to $^{13}$CO~$J=2-1$. We used the Forbes Filter Bank (FFB) backend in 4 IF mode, an upper and lower sideband each with 1~MHz and 250~kHz of spectral resolution, respectively. The channel width, $\Delta\nu_\mathrm{ch}$, of our spectrometer was $0.33$~km~s$^{-1}$. The 1~MHz resolution data were used to determine main beam efficiencies $\eta_\mathrm{mb}$, and the $250$~kHz resolution data were used to measure the $^{12}$CO line.

\begin{deluxetable}{lccc}
\tabletypesize{\scriptsize}
\tablewidth{0pt}
\tablecaption{Observation of Main Beam Efficiencies $\eta$\label{etatab}}
\tablehead{\colhead{Planet} & \colhead{$\eta_\mathrm{Vpol}$} & \colhead{$\eta_\mathrm{Hpol}$} & \colhead{$\eta_\mathrm{Vpol}/\eta_\mathrm{Hpol}$}}
\startdata
Mars\tablenotemark{a} & $0.80\pm0.05$ & $0.72\pm0.04$ & $1.10\pm0.09$\\
Mars & $0.80\pm0.05$ & $0.71\pm0.04$ & $1.12\pm0.09$\\
Saturn & $0.93\pm0.04$ & $0.70\pm0.03$ & $1.33\pm0.08$\\
Saturn & $0.94\pm0.04$ & $0.69\pm0.03$ & $1.36\pm0.09$\\
Venus & $0.89\pm0.03$ & $0.66\pm0.03$ & $1.35\pm0.08$\\
Venus & $0.88\pm0.03$ & $0.66\pm0.03$ & $1.34\pm0.08$\\
Mars & $0.87\pm0.05$ & $0.66\pm0.04$ & $1.31\pm0.10$\\
Mars\tablenotemark{b} & $0.87\pm0.05$ & $0.66\pm0.04$ & $1.33\pm0.10$\\
Mars\tablenotemark{c} & $0.88\pm0.05$ & $0.68\pm0.04$ & $1.29\pm0.10$\\
Mars & $0.89\pm0.05$ & $0.68\pm0.04$ & $1.31\pm0.11$\\
Mars & $0.84\pm0.04$ & $0.70\pm0.04$ & $1.21\pm0.10$\\
Mars & $0.88\pm0.05$ & $0.68\pm0.04$ & $1.29\pm0.10$\\
Venus & $0.93\pm0.03$ & $0.68\pm0.03$ & $1.38\pm0.09$
\enddata
\tablenotetext{a}{All Mars brightness temperature errors assumed to be 5\%}
\tablenotetext{b}{End of first night}
\tablenotetext{c}{Beginning of second night}
\end{deluxetable}

Using CLASS in the GILDAS data reduction package, we estimated the main beam efficiencies by observing the planets shown in Table~\ref{etatab}. Typical sideband rejections, ignored in the calibration, were $>10$~dB. The main beam efficiency $\eta$ was computed following \citet{mangum93} and corrected for single-sideband observations:
\begin{equation}
\label{etaequ}
\eta=\frac{T_A^{\star}(\mathrm{planet})}{J(\nu_s,T_\mathrm{planet})-J(\nu_s,T_\mathrm{cmb})}\times \left[1-\exp{\left(-\ln{(2)}\frac{\theta_\mathrm{eq}\theta_\mathrm{pol}}{\theta_\mathrm{mb}^2}\right)}\right]^{-1},
\end{equation}
where
\begin{equation}
\label{Jequ}
J(\nu,T_b)=\frac{h\nu/k}{e^\frac{h\nu}{k T_\mathrm{mb}}-1}
\end{equation}
is the Planck function at brightness temperature $T_b$ and frequency $\nu$, $T_A^\star$ is the single-sideband antenna temperature of the planet, $T_\mathrm{planet}$ is the planet's observed brightness temperature, $T_\mathrm{cmb} = 2.73$~K, $\theta_\mathrm{eq}$ and $\theta_\mathrm{pol}$ are respectively the planet's equatorial and poloidal diameters in arcseconds, and $\theta_\mathrm{mb}=33\arcsec$ at $\nu=230$~GHz. We adopted an average Venus brightness temperature $T_b$ from \citet{kuznetsov82} of $287\pm20$~K. For all other planets' $T_b$, we used the JCMT online database\footnote{\url{http://www.jach.hawaii.edu/jac-bin/planetflux.pl}}. We derived a ratio $\eta_\mathrm{Vpol}/\eta_\mathrm{Hpol}$ of the two IF's mean main beam efficiencies for both nights of $1.24\pm0.04$. We used this ratio to scale the Hpol polarization's antenna temperature up to match the level of the Vpol polarization's antenna temperature. After fitting a baseline to each spectrum, we averaged the sum of the scaled Hpol brightness temperatures and the Vpol brightness temperatures: $\frac{1}{2}\langle T_A^\star(\mathrm{Hpol, scaled})+T_A^\star(\mathrm{Vpol})\rangle=T_A^\star(\mathrm{sum})$. Thus we computed the corrected main beam temperature as
\begin{equation}
\label{tmb_equ}
T_\mathrm{mb}=\frac{T_A^\star(\mathrm{sum})}{\eta_\mathrm{Vpol}}.
\end{equation}
The average beam efficiencies were $\langle\eta_\mathrm{Hpol}\rangle=0.68\pm0.01$ and $\langle\eta_\mathrm{Vpol}\rangle=0.88\pm0.01$ for both nights.

Since we had null detections for our two sources, we must assume a line-width to calculate upper limits on the integrated intensity. We assumed typcial a line-width of $\Delta\nu=10$~km~s$^{-1}$ ($=7.69$~MHz). If we assume the CO line is well described by a Gaussian line shape, then the uncertainty in the integrated intensity is given by
\begin{equation}
\label{sigmai}
\sigma_I = \sigma_{T_\mathrm{mb}}\sqrt{\frac{3\Delta\nu_\mathrm{ch}\Delta\nu}{\sqrt{\ln{2}}}},
\end{equation}
where $\sigma_I$ and $\Delta v$ are the CO line fluxes and the the full width at half maximum (FWHM) and $\Delta v_\mathrm{ch}$ is the channel spacing 0.33~km~s$^{-1}$; see Appendix I of \citet{schlingman}. The observations are summarized in Table~\ref{obssummary}.

\begin{deluxetable}{lcccccc}
\tabletypesize{\scriptsize}
\tablewidth{0pt}
\rotate
\tablecaption{Observational Summary\label{obssummary}}
\tablehead{\colhead{Source} & \colhead{Integration Time} & \colhead{$\sigma_{T_\mathrm{mb}}$} & \colhead{$\sigma_I$} & \colhead{$\log_{10}{(F_{\nu})}$} & \colhead{$\overline{N}(T\mathrm{ex}=10,20,100\mathrm{~K})$} & \colhead{$M_{\mathrm{H}_2}(T_\mathrm{ex}=10,20,100\mathrm{~K})$} \\
\colhead{} & \colhead{(sec)} & \colhead{(K)} & \colhead{(K~km~s$^{-1}$)} & \colhead{($\log_{10}{(\mathrm{W~cm}^{-2})}$)} & \colhead{(10$^{13}$~cm$^{-2}$)} & \colhead{(M$_\earth$)}}
\startdata
RX~J0432.8+1735 & 3600 & 0.019 & 0.046 & $< -23.02$ & $<12.7,7.47,6.53$ & $<3.59,1.10,1.84$\\
St~34  & 6120 & 0.019 & 0.046 & $< -23.02$ & $<12.7,7.47,6.53$ & $<3.59,1.10,1.84$
\enddata
\tablecomments{$T_\mathrm{mb}$ is the main beam corrected brightness temperature and $I$ is the corresponding intensity assuming a line width of 5~km~s$^{-1}$. $F$ is the $3\sigma$ line flux upper limit.}
\tablecomments{That $\sigma_{T_\mathrm{mb}}$ for both objects is the same is a fluke.}
\end{deluxetable}

\section{Results \& Analysis}
\begin{figure}
\begin{center}
\plottwo{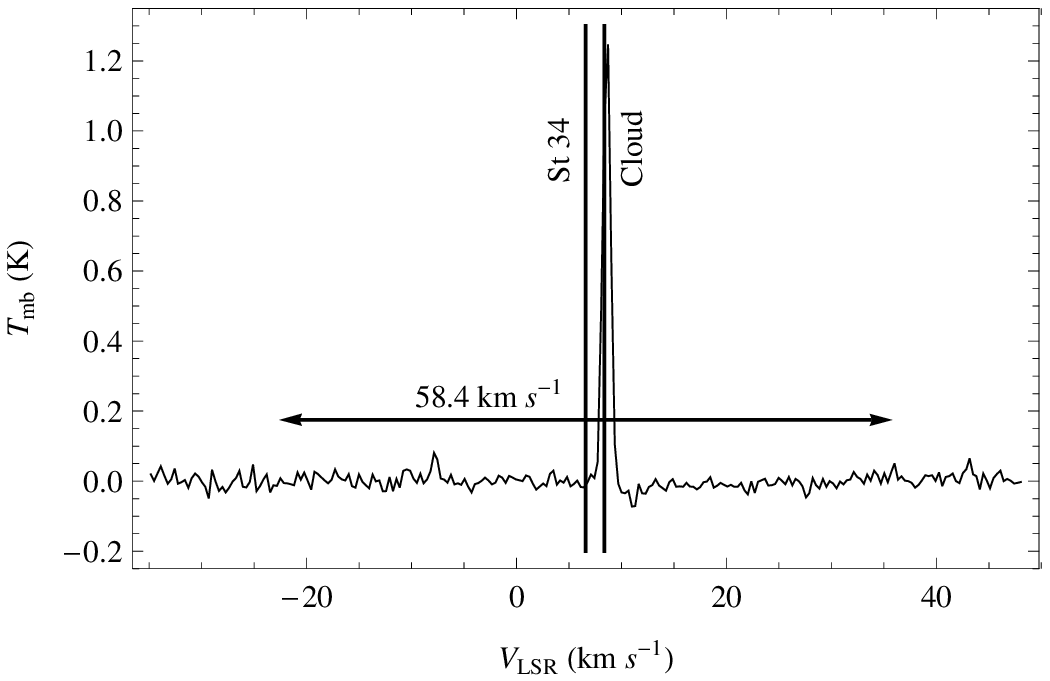}{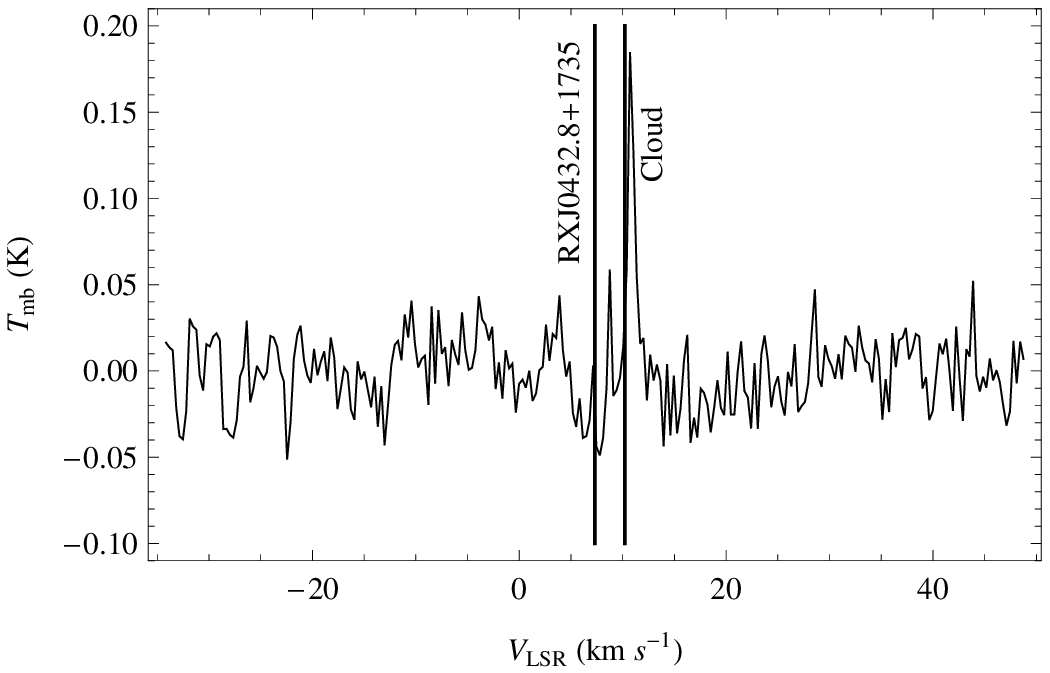}
\end{center}
\caption{Spectra of St~34 ({\it left}) and RX~J0432.8+1735 ({\it right}). The vertical line labeled ``Cloud" represents the estimated background cloud velocity determined by the \citet{dame87} CO~$J=1-0$ survey, and the vertical line labeled with the source's name indicates that source's $V_\mathrm{LSR}$. For St~34, we show with an arrow the difference between the radial velocities of its binary components with an arrow centered on the systemic velocity.}
\label{spec}
\end{figure}

The main-beam corrected spectra of our observations are shown in Figure~\ref{spec}.

While 9$\arcmin$ is a rough spatial scale for comparison to the 33$\arcsec$ beam of the SMT, and since we did not detect a $^{12}$CO line in any of our sources, the on-cloud results from SMT are consistent with \citet{dame87}. It is unlikely that high spatial frequency variations of 33$\arcsec$ scales over 9$\arcmin$ regions have systemic velocity shifts of 2-3~km$^{-1}$ 

Since we did not detect any $^{12}$CO line, we convert our $3\sigma$ noise into an upper limit on the flux. A knowledge of the flux will enable us to estimate upper limits on gas disk mass.

The observed flux is the double-integral of the observed intensity $I_\nu$ over frequency $\nu$ and solid angle $\Omega$:
\begin{equation}
\label{double integral}
F=\int\int I_\nu\,d\nu\,d\Omega=\int\int\frac{2 k \nu^3 T_b}{c^3}\,dv\,d\Omega.
\end{equation}
Assuming the brightness temperature does not vary substantially over the telescope beam and that the line is a Gaussian with line-width $\Delta v$, then the upper limit on the $3\sigma$ line flux $F$ is
\begin{equation}
\label{upper lim}
F < \frac{2 k \nu^3 (3\sigma_{T_\mathrm{mb}})}{c^3}\frac{\pi\theta^2}{4\ln{2}}\sqrt{\frac{4\ln{2}}{\pi}}\times\Delta v=(5.05\times10^{-15})\sigma_{T_\mathrm{mb}}\mathrm{erg~s}^{-1}\mathrm{~cm}^{-2}.
\end{equation}
The $3\sigma$ upper limits on $F$ are listed in Table~\ref{obssummary}.

Similarly, we can derive the column density in the optically thin limit to be
\begin{equation}
\label{col depth equ}
\overline{N}(T_\mathrm{ex})=\frac{8\pi k\nu^2}{h c^3 g_u A_{ul}}\mathcal{F}(T_\mathrm{ex}, E_u, \nu)\int T_\mathrm{mb}\,dv,
\end{equation}
where
\begin{equation}
\label{F def}
\mathcal{F}(T_\mathrm{ex}, E_u, \nu)\equiv\frac{J_\nu(T_\mathrm{ex})Q(T_\mathrm{ex})\exp{\left(\frac{E_u}{k T_\mathrm{ex}}\right)}}{J_\nu(T_\mathrm{ex})-J_\nu(T_\mathrm{cmb})},
\end{equation}
$A_{ul}$ is the Einstein A coefficient (spontaneous emission) and has units of s$^{-1}$.

Similar to the analysis of \citet{pascucci06}, we assume an excitation temperature $T_\mathrm{ex}\approx20$~K. Then in our case for $^{12}$CO~$J=2-1$, $\mathcal{F}(T_\mathrm{ex}, E_u, \nu)\approx28.00$. The Einstein $A_{ul}=6.91\times10^{-7}$~s$^{-1}$ and partition function $Q(20\mathrm{~K})=15.9$ (CDMS\footnote{\url{http://www.ph1.uni-koeln.de/vorhersagen/}}).

We compute and tabulate in Table~\ref{obssummary} the $^{12}$CO number densities and H$_2$ gas masses in the optically thin limit. Gas disk masses were derived from \citet{scoville86},
\begin{equation}
\label{scoville mass eqn}
M_{\mathrm{H}_2}<\overline{N}(T_\mathrm{ex})\times\Bigg\{\left[\frac{\mathrm{H}_2}{\mathrm{CO}}\right]\mu_G m_{\mathrm{H}_2} \frac{\pi\theta^2}{4} d^2\Bigg\}\mathrm{~g}=110\sigma_{T_\mathrm{mb}}\mathrm{~M}_\earth,
\end{equation}
where $\left[\mathrm{H}_2/\mathrm{CO}\right]\approx10^4$, $\mu_G=1.36$ is the mean molecular weight, $m_{\mathrm{H}_2}$ is the mass of an H$_2$ molecule, and $d\approx140$~pc is the distance to Taurus.




\section{Discussion}

Theories of gaseous planet formation require knowledge of (1) how much gas there is in circumstellar disks initially and (2) the rate at which gas is depleted over time. To answer the first question, upper limits on the amount of gas in very young protostellar disk systems puts limits directly on the amount of mass available for gaseous planet formation in the systems observed. Answers to the second question, requiring large samples over a wide range of ages, are also necessary because of the competition between the timescale for planet formation and gas disk dispersion timescales \citep[for a recent review, see][]{meyer09}.


There are several ways of constraining gas disk masses, each with its own advantages and disadvantages. To understand gas disk timescales, one can also analyze H$\alpha$ emission line profiles and determine gas accretion rates
and thus constrain the gas mass surface density at the inner edge of the disk tracing perhaps global disk evolution \citep[e.g.][]{fedele09}. Using UV tracers of gas emission, \citet{ingleby09} describe {\it HST} observations searching 
for evidence of hot gas in emission finding no evidence for H$_2$ emission for WTTS in their sample. Ultraviolet absorption line from a continuum source \citep[e.g.][]{roberge05} can help constrain cold mass in disks, but it requires a continuum source that is bright in the far UV and an edge-on geometry; therefore, it is observationally feasible only in special circumstances. 
Near IR fluorescent H$_2$ traces gas with excitation temperatures $T_\mathrm{ex}>2000$~K \citep{bary03} and mid-IR ro-vibrational lines \citep[e.g.][]{najita07,thi01,pascucci06}, especially the $28.2~\mu$m and $17\mu$m {\it Spitzer} bands, trace gas up to 50-200~K. Our observations of rotational lines of CO trace cooler gas at larger orbital radii. The disadvantage of such measurements is that CO can freeze out at the coldest temperatures corresponding to the outer limits of the disk $\gtrsim 30$-100~AU. \citet{hughes10} present evidence for evolving gas to dust ratios in transitional disks which exhibit evidence for optically-thick outer disks but possess inner holes and gaps.



St~34, being $8\pm3$~Myr old, might have a lower gas surface density than typical CTTSs. It is still accreting gas \citep{white05}, albeit at a low rate, but retains at least a low density inner disk. Our non--detection in a search for cold gas implies that most of its outer disk must have disappeared or frozen out onto grains. \citet{white05} show that St~34 has an accretion rate $\langle\dot{M}\rangle=8.3\times10^{-10}$~M$_\sun$~yr$^{-1}$, so after 1~Myr much more gas than our upper limit of $4.20$~M$_\earth$ would accrete ($\sim276$~M$_\earth$). Perhaps St~34 recently lost its outer disk through photoevaporation \citep[e.g.][]{gorti09} and we are witnessing the ``last gasp'' of accretion onto the star. This is possible but not likely, as it requires current observations to be taking place at a very special time.

Conversely, RX~J0432.8+1735---being a much younger, $1.0\pm0.5$~Myr WTTS---must have either evaporated its disk or formed planets from its gas disk faster than normal. Its being a WTTS is consistent with our null detection of its gas content, although remnant amounts of gas less than a few $M_\earth$ could still exist in its outer disk. 
That RX~J0432.8+1735 is relatively young and does not have detectable gas content could be significant considering that there are older systems, such as Hen 3-600, a binary system at between $1-10$~Myr of age with apparent WTTS and CTTS components \citep{jayawardhana99}; TW~Hya, a CTTS at 8-10~Myr \citep{webb99}; and DM~Tau at $\sim 8$~Myr \citep{guilloteau94}. Thus something about the disk evaporation physics is dramatically different in RX~J0432.8+1735 than these oldest systems.

\begin{figure}
\begin{center}
\plotone{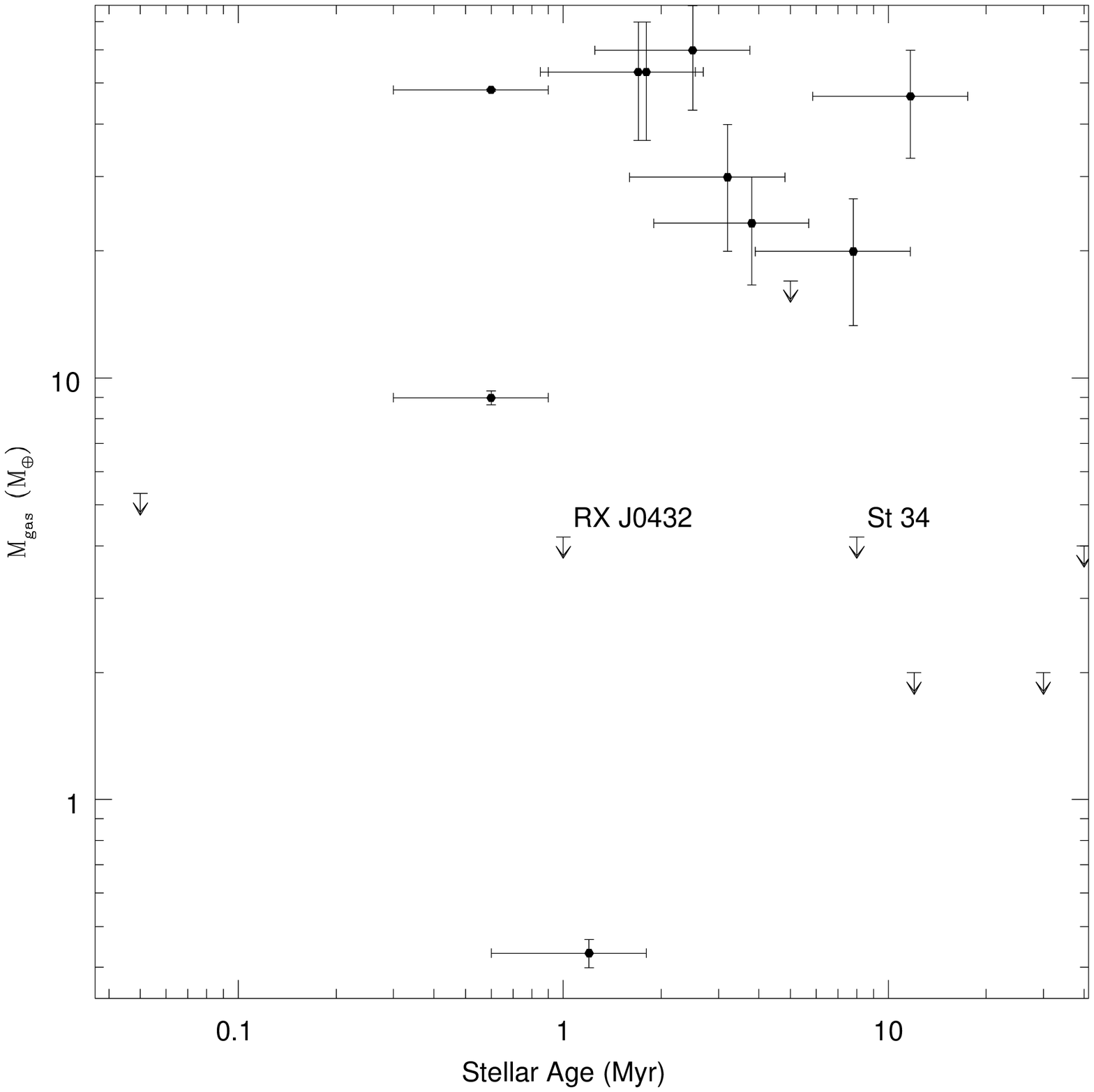}
\end{center}
\caption{Gas circumstellar disk mass versus age of selected sources: our RX~J0432.8+1735 and St~34 upper limits ({\it labeled}); upper limits from the \citet{pascucci06} sample ({\it unlabeled upper limits}) with ages $\le 30$~Myr (the Kelvin-Hemholtz contraction timescale for a 1~M$_\sun$ star); and exact mass determinations ({\it solid circles}) of BP~Tau \citep[$^{13}$CO~$J=2-1$;][]{dutrey03}; DL~Tau, DO~Tau \citep[$^{12}$CO~$J=2-1$;][]{koerner95}; and DM~Tau, DR~Tau, GG~Tau~a, GM~Aur, GO~Tau, LkCa~15, RY~Tau \citep[$^{12}$CO~$J=3-2$ and $^{13}$CO~$J=3-2$;][]{thi01}. We assume errors in stellar ages to be 50\%.}
\label{mgdisk vs age}
\end{figure}

In Figure~\ref{mgdisk vs age} we compare our sources' gas disk masses and ages with the gas mass upper limits of those sources $\le 30$~Myr from \citet{pascucci06} and with the gas mass determinations ({\it solid circles}) of BP~Tau \citep[$^{13}$CO~$J=2-1$;][]{dutrey03}; DL~Tau, DO~Tau \citep[$^{12}$CO~$J=2-1$;][]{koerner95}; and DM~Tau, DR~Tau, GG~Tau~a, GM~Aur, GO~Tau, LkCa~15, RY~Tau \citep[$^{12}$CO~$J=3-2$ and $^{13}$CO~$J=3-2$;][]{thi01}. We note that this is not an exhaustive compilation from the literature, but representative of recent results. Assuming a 1:10 gas-to-dust ratio \citep{dalessio05}, we would expect RX~J0432.8+1735 and St~34 to have at most 0.420~M$_\earth$ and 0.420~M$_\earth$ of dust, respectively, for $T_\mathrm{ex}=20$~K. For St~34, \citet{hartmann05} estimates a disk mass of 665~M$_\earth$ located in a circumbinary disks between the ``wall" (the region defined to surround the two components of the St~34 binary; $\sim 0.7$~AU) and $7$~AU. Assuming this mass is representative of a total disk mass in St~34 out to $7$~AU, this would be consistent with a small ($<10$~AU), optically-thick CO disk with a gas-to-dust ratio of $\sim100$; we do not detect this due to beam dilation \citep[cf.][]{pascucci06}. 




St~34 has infrared excess for wavelengths longer than 3.6~$\mu$m \citep{hartmann05} and RX~J0432.8+1735 has infrared excess for wavelengths longer than 24~$\mu$m \citep{padgett06}, yet St~34 is a binary CTTS with an accreting inner disk and RX~J0432.8+1735 is a WTTS. Binaries tend to disrupt inner gas disks \citep{jensen96} and may decrease disk lifetimes \citep{monin07}. However, \citet{armitage96} have argued that close binaries affect angular momentum exchange in the natural evolution of accretion disks resulting in longer lived outer disks. Indeed \citet{thebault04} find that planet formation around binaries might require a long-lived but massive disk. Since circumbinary disks allow for long gas disk lifetimes, St~34 might have had more time to form planets.


\section{Conclusions}

Assuming optically thin disks ($\tau\ll1$), an excitation temperature $T_\mathrm{ex}=20$~K, and a line-width $\Delta v=10$~km~s$^{-1}$, we do not detect significant amounts of gas around three T~Tauri stars: $<4.20$~M$_\earth$ for the PMS binary St~34 and $<4.20$~M$_\earth$ for RX~J0432.8+1735. St~34, a CTTS, is still accreting gas although it is $8\pm 3$~Myr old, and the gas disk of RX~J0432.8+1735, a WTTS of $1.0\pm0.5$~Myr, has disappeared quickly. 
Future observations of larger samples will be required to understand the diversity of disk lifetimes as a function of stellar properties.

\end{document}